\documentclass[aps,prl,twocolumn,showpacs,superscriptaddress,superscriptaddress,
longbibliography,msuperscriptaddress]{revtex4-1}
\usepackage[english]{babel}
\usepackage[pdftex]{graphicx}
\usepackage{geometry}
\usepackage{graphicx}
\usepackage{hyperref}
\usepackage{floatrow}
\usepackage{subfigure}
\usepackage{floatflt}
\usepackage{amsmath}
\usepackage{floatflt,graphicx}
\usepackage{braket}
\usepackage{xcolor}
\usepackage{amssymb}
\usepackage{amsmath,amssymb,amsthm,mathrsfs,amsfonts,dsfont}
\usepackage{parskip}
\usepackage{bbold}
\usepackage{soul}

\begin{document}

\title[]{Entanglement Swapping with Semiconductor-generated Photons}

\author{Michael~\surname{Zopf}}
\affiliation{Institute for Integrative Nanosciences, Leibniz IFW Dresden, Helmholtzstra{\ss}e~20, 01069~Dresden, Germany}
\author{Robert~\surname{Keil}}
\affiliation{Institute for Integrative Nanosciences, Leibniz IFW Dresden, Helmholtzstra{\ss}e~20, 01069~Dresden, Germany}
\author{Yan~\surname{Chen}}
\affiliation{Institute for Integrative Nanosciences, Leibniz IFW Dresden, Helmholtzstra{\ss}e~20, 01069~Dresden, Germany}
\author{Jingzhong~\surname{Yang}}
\affiliation{Institute for Integrative Nanosciences, Leibniz IFW Dresden, Helmholtzstra{\ss}e~20, 01069~Dresden, Germany}
\affiliation{Institut f{\"u}r Festk{\"o}rperphysik, Leibniz Universit{\"a}t Hannover, Appelstra{\ss}e~2, 30167~Hannover, Germany}
\author{Disheng~\surname{Chen}}
\affiliation{Institute for Integrative Nanosciences, Leibniz IFW Dresden, Helmholtzstra{\ss}e~20, 01069~Dresden, Germany}
\author{Fei~\surname{Ding}}\email{f.ding@fkp.uni-hannover.de}
\affiliation{Institute for Integrative Nanosciences, Leibniz IFW Dresden, Helmholtzstra{\ss}e~20, 01069~Dresden, Germany}
\affiliation{Institut f{\"u}r Festk{\"o}rperphysik, Leibniz Universit{\"a}t Hannover, Appelstra{\ss}e~2, 30167~Hannover, Germany}
\author{Oliver G.~\surname{Schmidt}}\email{o.schmidt@ifw-dresden.de}
\affiliation{Institute for Integrative Nanosciences, Leibniz IFW Dresden, Helmholtzstra{\ss}e~20, 01069~Dresden, Germany}
\affiliation{Material Systems for Nanoelectronics, Technische Universit{\"a}t Chemnitz, 09107~Chemnitz, Germany}

\begin{abstract}
Transferring entangled states between photon pairs is essential for quantum communication technologies. Semiconductor quantum dots are the most promising candidate for generating polarization-entangled photons deterministically. Recent improvements in photonic quality and brightness now make them suited for complex quantum optical purposes in practical devices. Here we demonstrate for the first time swapping of entangled states between two pairs of photons emitted by a single quantum dot. A joint Bell measurement heralds the successful generation of the Bell state $\Psi^+$  with  a fidelity of up to $0.81 \pm 0.04$. The state's nonlocal nature is confirmed by violating the CHSH-Bell inequality. Our photon source is compatible with atom-based quantum memories, enabling implementation of hybrid quantum repeaters. This experiment thus is a major step forward for semiconductor based quantum communication technologies.
\end{abstract}

\maketitle

Semiconductor light sources have revolutionized science and technology since laser diodes~\cite{Hall1962,Kroemer1963} and vertical-cavity surface-emitting lasers (VCSELs)~\cite{Melngailis1965,Soda1979} arrived in the 1960's. Quantum mechanics lies at the roots for these devices, yet quantum states of light have only in recent decades been studied extensively in their own right - sparking the "second quantum revolution". Semiconductor sources can now emit single photons~\cite{Somaschi2016} and entangled photons~\cite{Salter2010} on demand (see Fig. \ref{fig1}a), more reliably and intensely than non-linear crystals. They hold great potential for a range of applications in quantum communication~\cite{Gisin2002}, quantum metrology~\cite{Dowling2008} and quantum computation~\cite{Knill2001}.

The next step towards building quantum networks is to transfer entangled states between distinct pairs of photons~\cite{Kimble2008,Bose1998,Briegel1998}. This entails substituting the pairwise entanglement in two-photon states with entanglement between photons from different pairs~\cite{Bennett1993,Zukowski1993}. The first experiment to do this two decades ago~\cite{Pan1998} used a technique based on spontaneous parametric down conversion in a nonlinear optical crystal~\cite{Kwiat1995,Shih1988}. Though such sources are widely used, for example to entangle multiple photons~\cite{Wang2016}, their brightness and therefore scalability is fundamentally limited owing to Poissonian emission statistics~\cite{Scarani2005}.

Semiconductor quantum dots (QDs), by contrast, are able to generate entangled photon pairs deterministically one by one~\cite{Akopian2006}. However, until recently, QDs were too faint and of poor degree of entanglement and indistinguishability to use for advanced quantum applications. Improvements of the past three years have overcome these limitations. Highly coherent~\cite{Jahn2015} and strongly entangled photons~\cite{Keil2017,Huber2017} can now be generated with high brightness~\cite{Chen2018} and reproducibility~\cite{Keil2017} from QDs.

Here we demonstrate, for the first time, entanglement swapping between polarization-entangled photons emitted by a semiconductor QD. The Bell state $\Psi^+$ is generated with high fidelity and strong nonlocal characteristics, proven by violating the CHSH-Bell inequality~\cite{Bell1964,CHSH1969}. Our semiconductor sources are compatible with atom-based quantum memories. This opens up their use in devices such as quantum repeaters (the quantum equivalent of a classical amplifier)~\cite{Rakher2013} which are essential for long distance quantum communication.

\section{Entangled state generation}
\begin{figure}[t]
	\includegraphics[width=1\textwidth]{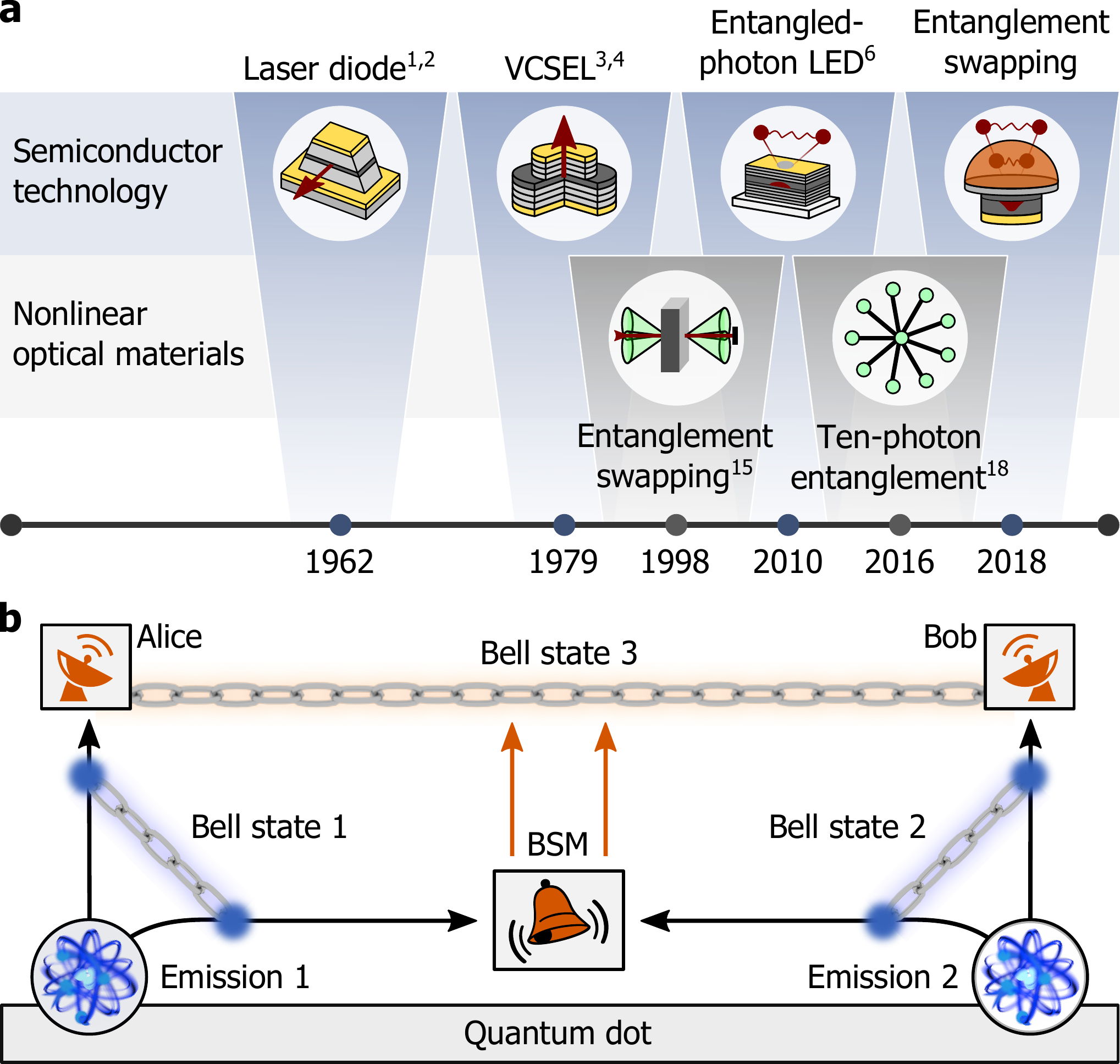}
	\caption{ \textbf{(a)}
	Historical development of integrated entangled photon sources, starting from semiconductor lasers and photonic entanglement based on nonlinear optical materials, to scalable quantum dot sources of entangled photons.
	\textbf{(b)} Principle of an entanglement swapping experiment using a quantum dot. Two distinct pairs of entangled photons are generated (emission 1 and 2). One photon from each pair is directed to a Bell state measurement (BSM). Upon success, the BSM establishes entanglement of the remaining photons sent to Alice and Bob.}
\label{fig1}
\end{figure}

\begin{figure*}[ht]
	\includegraphics[width=1\textwidth]{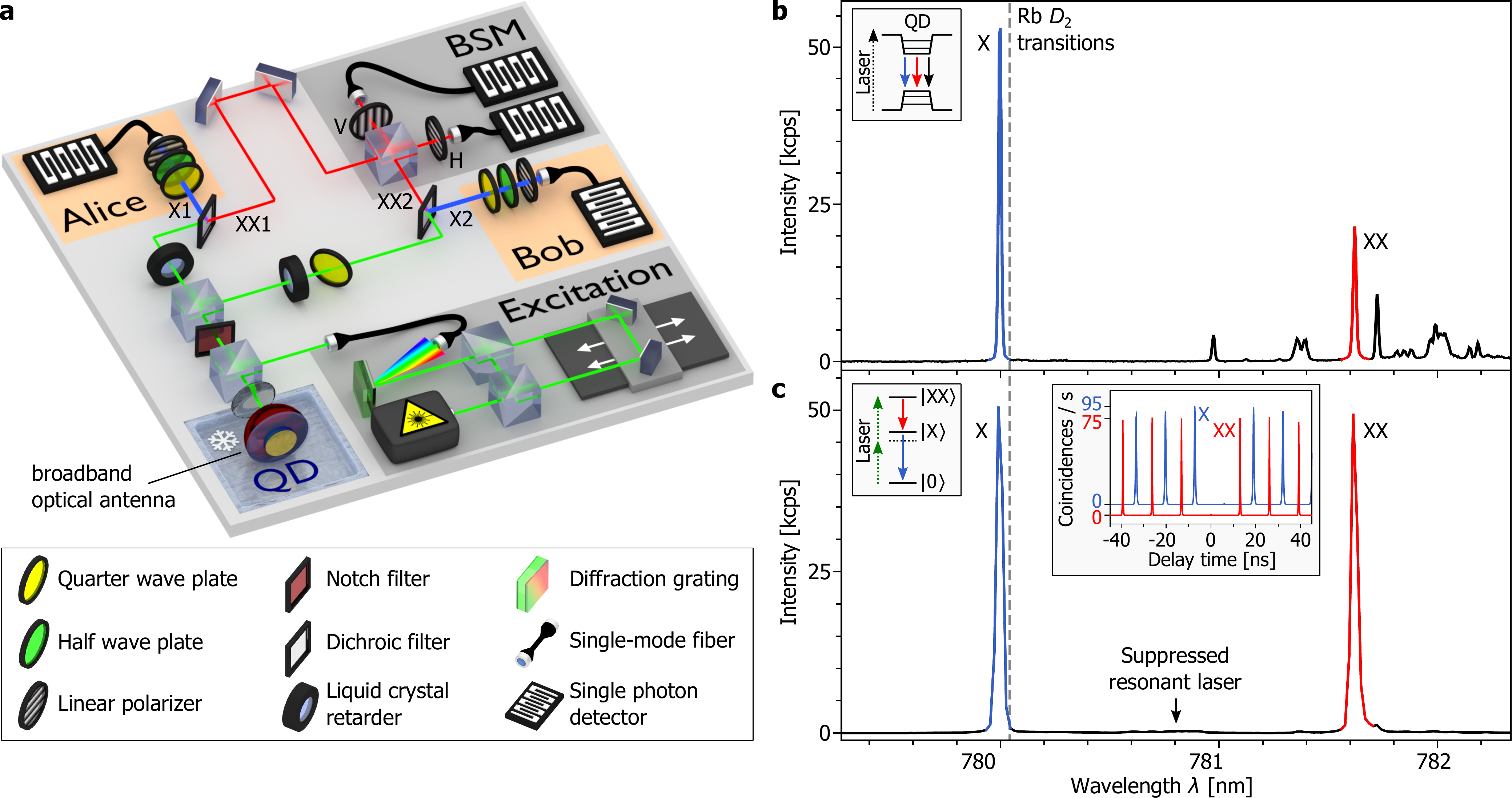}%
	\caption{\textbf{Experimental setup and quantum dot emission spectra. (a)} Entanglement swapping setup. Two consecutive pairs of polarization-entangled photons Xi--XXi (emission i = 1,2) are generated by resonantly exciting a quantum dot (QD) embedded in an optical antenna. The emitted light is cleansed of residual laser signal and then sent to a non-polarizing beam splitter. Two photons XX1 and XX2 from each emission are directed to a Bell state measurement (BSM). Coincidence detection heralds the polarization-entanglement of the remaining photons X1 and X2. The latter are guided to two polarization-analyzers Alice and Bob.
\textbf{(b)} QD photoluminescence spectrum under above-bandgap excitation highlighting the most prominent features, the exciton (X) and bi-exciton (XX) emissions. \textbf{(c)} Emission spectrum obtained by pulsed resonant two-photon excitation of the bi-exciton state. Decay via the intermediate exciton states results in the emission of spectrally distinct, polarization-entangled photon pairs XX--X. The inset shows the intensity autocorrelation for each spectral feature, indicating a high single-photon purity of $g^{(2)}(0) \leq 0.005$.}
	\label{fig2}
\end{figure*}

The experimental concept is sketched in Fig.~\ref{fig1}b. Two pairs of polarization-entangled photons are consecutively emitted (emission 1 and 2) by a single semiconductor quantum dot. The polarization of one photon from each pair is measured by separate detectors, labeled Alice and Bob. Then, a joint Bell state measurement (BSM) is made on the remaining two photons; this swaps the entanglement of the original pairs to the photons that Alice and Bob receive.  The source of entangled photons in our experiment are GaAs/AlGaAs QDs grown by local droplet etching, as they are reliable and reproducable to make with entanglement fidelities close to unity~\cite{Keil2017} and highly indisitinguishable photons~\cite{Huber2017}. The QDs are embedded in a nanomembrane which is sandwiched by a silver reflector and a spacing layer, attached to a gallium phosphide hemsipherical lens~\cite{Chen2018}. This design provides a broadband optical antenna, offering photon extraction efficiencies up to 65\,\% while preserving a high single photon purity and entanglement fidelity.\\

The QD antenna's operating temperature of $T=4$\,K is reached using a closed-cycle helium cryostat. A selected QD is first triggered by optically pumping the surrounding host semiconductor material. The emission spectrum in Fig.~\ref{fig2}b displays two prominent features: the exciton (X) emission at $780.0$\,nm and the bi-exciton (XX) emission at $781.6$\,nm. Here, the X photons reside near the optical $\mathrm{D_2}$ transitions of rubidium, a prominent quantum memory candidate~\cite{Koerber2018}.

To generate two consecutive polarization-entangled photon pairs (emission 1 and 2), we exploit the bi-exciton(XX)-exciton(X) radiative cascade~\cite{Akopian2006}. Deterministic excitation of the XX state is ensured by resonant two-photon excitation~\cite{MullerM2014}. A pair of photons is emitted in the successive decay via the intermediate X states to the ground state (left inset of Fig.~\ref{fig2}c). The photons share the polarization-entangled Bell state $\ket{\Phi^+}_{i}$ in the respective emission $i=1,2$:
\begin{equation}
\ket{\Phi^+}_{i} = \ket{H_XH_{XX}} + \ket{V_XV_{XX}}.
\end{equation}
with $H$ and $V$ representing horizontal and vertical polarization of the rectilinear basis.

 For efficient resonant excitation we use a pulsed Ti:sapphire laser operating at a $76$MHz repetition rate. Guiding the laser light into a tunable, unbalanced Mach-Zehnder interferometer yields two consecutive excitation pulses. The laser's spectral width is reduced and the central wavelength adjusted by a successive diffraction grating and single-mode fiber. Thus the laser emission wavelength is fixed at the XX two-photon resonance between the X and XX emission. Notch filters are used to suppress the scattered laser in the QD emission signal. The signal intensity is enhanced further by exciting the QD by a weak continuous wave laser emitting at 650nm.

Fig.~\ref{fig2}c shows the resulting emission spectrum of the XX cascade emission and a well-suppressed resonant laser. The right inset shows the intensity auto-correlation $g^{(2)}(\tau)$ of the X and XX emissions obtained in a Hanbury Brown and Twiss measurement~\cite{HBT1956}. Vanishing coincidences at zero delay time bear witness to a high single photon purity, with values of $g^{(2)}_{X}(0)=0.0041 \pm 0.0003$ and $g^{(2)}_{XX}(0)=0.0050 \pm 0.0005$. We attribute the residual coincidences at zero delay to laser background, which can in principle be suppressed further using additional notch filters.

So far the two photon pairs share the same light path upon emission. As laid out in Fig.~\ref{fig2}a, a non-polarizing beam splitter is used together with time-gated detection in order to separate emissions 1 and 2. The XX and X photons from each pair are split apart using dichroic optical filters, which transmit XX photons and reflect the X emission.

At this stage, the four-photon state $\ket{\alpha}$ is a product of the states from emissions 1 and 2. It can be rewritten into products of Bell states between the X and XX photons:
\begin{equation}
\begin{split}
\ket{\alpha} &= \ket{\Phi^+}_1  \ket{\Phi^+}_2\\
 &= \frac{1}{2}( \ket{\Phi^+}_{X} \ket{\Phi^+}_{XX} + \ket{\Phi^-}_{X} \ket{\Phi^-}_{XX}\\
&\ \ \ \ \ \ \ + \ket{\Psi^+}_{X} \ket{\Psi^+}_{XX} + \ket{\Psi^-}_{X} \ket{\Psi^-}_{XX} )
\end{split}
\end{equation}
with the four polarization Bell states being
\begin{equation}
\begin{split}
\ket{\Phi^{\pm}} &= \ket{HH} \pm \ket{VV}\\
\ket{\Psi^{\pm}} &= \ket{HV} \pm \ket{VH}
\end{split}
\end{equation}
Projecting $\ket{\alpha}$ to a Bell state between photons XX1 and XX2 will in turn result in a Bell state shared by the previously uncorrelated X1 and X2. We project to the state $\ket{\Psi^{+}}$ by performing the following BSM: First, photons XX1 and XX2 are sent to interfere on a non-polarizing beam splitter. To ensure successful quantum interference, the arrival times of XX1 and XX2 have to be matched. Therefore the XX1 photons are delayed before the BSM, in order to compensate for the time difference between emission 1 and 2. After interference, the photons pass through an H- or V-oriented polarizer in each beam splitter output, respectively. Single-mode fibers then deliver the photons to superconducting nanowire single photon detectors (SNSPDs) with time resolutions of $50$ps.\\

Successful coincidence detection at the BSM now leaves the two remaining photons X1 and X2 in the Bell state
\begin{equation}
\ket{\Phi^{+}}_{AB} = \ket{HV} + \ket{VH}
\end{equation}
sent to Alice and Bob for measurement. Subsequent arrangement of a quarter-wave plate, half-wave plate, polarizer and SNSPD allows for projection on any desired polarization state.
In order to compensate for an accumulated phase and retardation in the setup we employ liquid crystal retarders and tilted quarter wave plates.

\section{Initial state characterization}
\begin{figure}[ht]
	\includegraphics[width=1\textwidth]{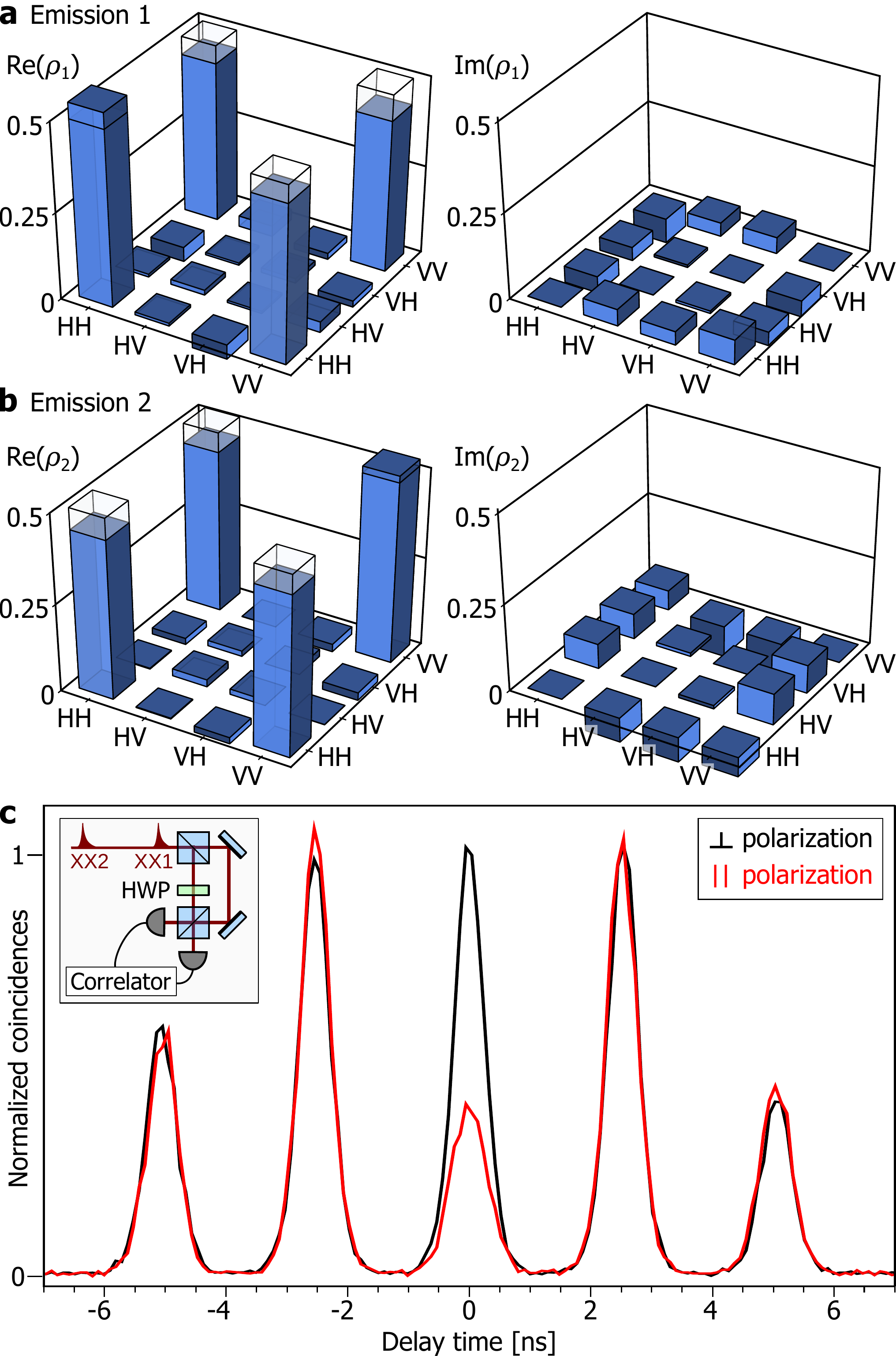}
	\caption{\textbf{Degree of entanglement and photon indistinguishability.} Two-photon density matrices of the photon pairs Xi--XXi from \textbf{(a)} emission i=1 and \textbf{(b)} emission i=2. Real (left) and imaginary part (right) closely resemble the Bell state $\ket{\Phi^+}$ with fidelities of $f_1=0.9369\pm0.0004$ (emission 1) and $f_2=0.9267\pm0.0004$ (emission 2). The shaded areas represent the difference to the ideally obtainable values.
\textbf{(c)} The indistinguishability $I=0.569 \pm 0.009$ of photons XX1 and XX2 is derived from a Hong-Ou-Mandel measurement. Coincidences are detected at the output of an unbalanced Mach-Zehnder interferometer (inset) and binned according to their detection delay times. Using a half-wave plate (HWP), co-polarized photons yield reduced coincidences (red) compared with crossed polarizations (black).}
	\label{fig3}
\end{figure}

Successful entanglement swapping relies on high entanglement fidelities $f_i$ of the initial photon pairs (emission $i=1,2$) and on high photon indistinguishabilities $I$ of the XX photons sent to the BSM. We perform quantum state tomography~\cite{James2001} to reconstruct the full two-photon density matrix $\rho_i$ of emissions $i=1,2$ as shown in Fig.~\ref{fig3}a and Fig.~\ref{fig3}b, respectively. The real (left) and imaginary parts (right) clearly resemble the Bell state $\ket{\Phi^+}$. We obtain fidelities of $f_1=0.9369\pm0.0004$ (emission 1) and $f_2=0.9267\pm0.0004$ (emission 2) indicating highly entangled photon emission. We attribute the deviation from unity fidelity to the slightly polarizing optical filters in the setup and the minor presence of a QD emission in the red-shifted vicinity of the XX photons (see Fig.~\ref{fig2}c). The marginally lifted exciton spin degeneracy by $(0.4\pm0.1)\,\mathrm\mu$eV, evanescent laser light background and polarization-dephasing during the QD's emission process are expected to have only a small effect on the fidelity~\cite{Keil2017}.\\

\begin{figure*}[ht]
	\includegraphics[width=1\textwidth]{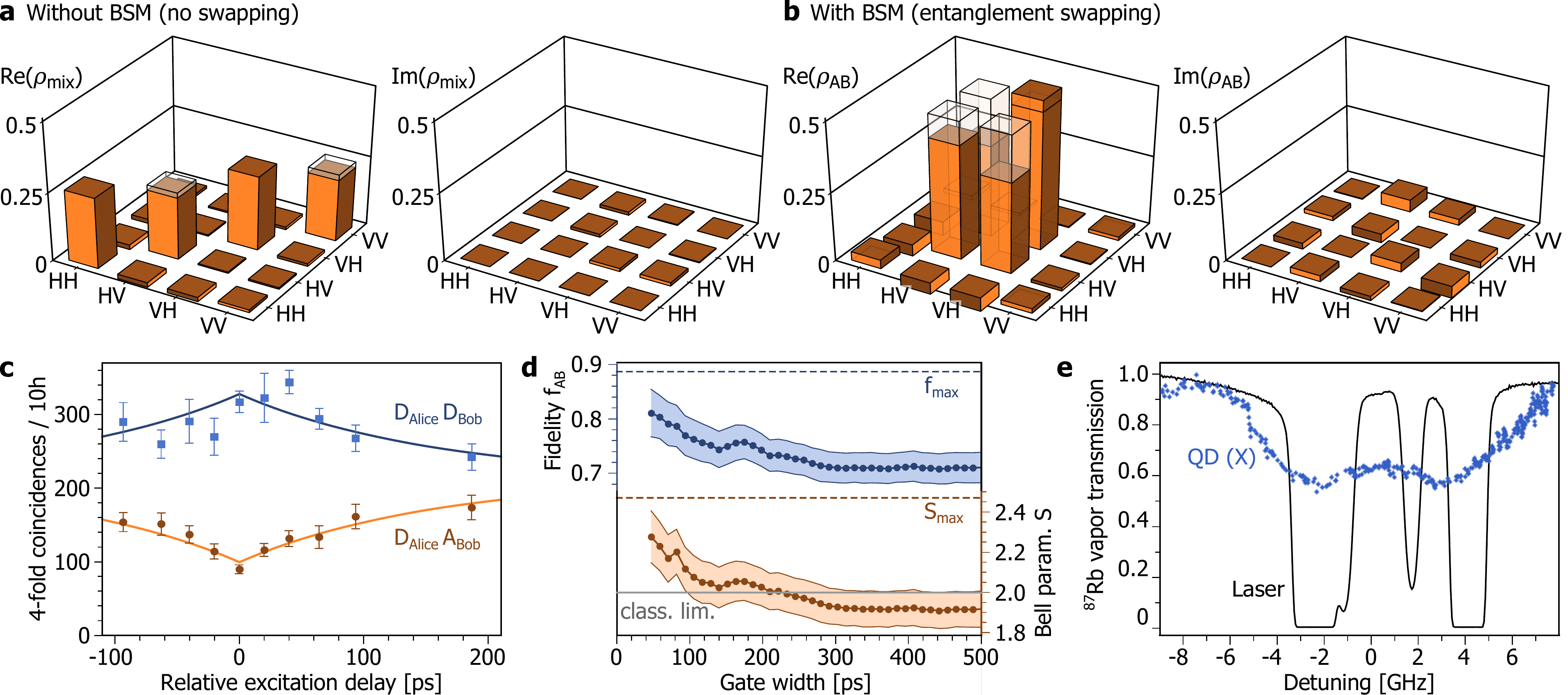}
	\caption{\textbf{Entanglement swapping with semiconductor-generated photons.} Density matrix of the two-photon state received by Alice and Bob without \textbf{(a)} and with \textbf{(b)} a heralding Bell state measurement (BSM). The shaded areas represent the difference to the ideally obtainable values. Real part (left) and imaginary part (right) of $\rho_{mix}$ show the distinct signature of a perfect statistical mixture $\frac{1}{4}\mathbb{1}$, whereas $\rho_{AB}$ closely resembles the entangled state  $\ket{\psi^+}$ with a fidelity of $f_{AB}=0.81\pm0.04$.
	\textbf{(c)} Four-fold coincidences as a function of the delay between photons XX1 and XX2 at the BSM setup. Measurement settings of Alice and Bob in the co-polarized (orange) and cross-polarized (blue) diagonal bases reveal a large difference at zero time delay, as expected in an entanglement swapping experiment. The solid lines denote the double-sided exponential fit.
	\textbf{(d)} Fidelity $f$ and Bell parameter $S$ as a function of gate width of photon detection at the BSM. Large gate widths result in a decreased fidelity of $f_{AB} = 0.71 \pm 0.03$. At $47$\,ps gate width, $S=2.28 \pm 0.13$ is obtained, violating the CHSH-Bell inequality. The dotted lines are the maximally achievable values in case of perfect photon indistinguishability.
	\textbf{(e)} $^{87}$Rb vapor cell transmission over the frequency detuning at the $D_2$ transitions for a narrow laser and the X photons from the temperature-tuned QD. Two absorption features are visible in the QD emission, enabling possible atom-semiconductor based quantum repeater applications.
	}
	\label{fig4}
\end{figure*}
Fig.~\ref{fig3}c shows a coincidence histogram obtained in an indistinguishability measurement~\cite{Santori2002} based on Hong-Ou-Mandel interference~\cite{Hong1987}. The two consecutive XX photons are guided into an unbalanced Mach-Zehnder interferometer featuring a time delay identical to that between XX1 and XX2. Indistinguishable XX photons will interfere at the second beam splitter and exit it pairwise, observable in the detection of reduced photon coincidences. Using a half-wave plate (HWP), the photon polarizations at the beam splitter can be made orthogonal. This renders the photons distinguishable which in turn gives rise to coincidences. At zero delay time between the detection events, the coincidences for parallel polarizations (red) show a significant reduction in comparison with those for perpendicular polarizations (black). We extract photon indistinguishabilities of $I=0.569 \pm 0.009$, which directly specifies the success probability of the BSM in the entanglement swapping experiment. The offset from unity arises most likely from internal dephasing processes and spectral jittering. Further spectral filtering or time-gating of detection events in the BSM can circumvent these effects at the expense of the BSM coincidence rate.

\section{Entanglement Swapping}
Having ensured high-fidelity emission of entangled photons we can focus on the execution of the entanglement swapping experiment. As a control measurement, the photon state shared by Alice and Bob is first investigated without considering the BSM. The density matrix $\rho_{mix}$ extracted from our observations via quantum state tomography is shown in Fig.~\ref{fig4}a. The signature of a statistical mixture of polarization states is evident, with a fidelity of $f_{mix}=0.9960 \pm 0.0004$ to the completely mixed state $\frac{1}{4}\mathbb{1}$. This is expected, since the photons X1 and X2 do not stem from the same emission cascade.\\

Now the entanglement shall be swapped from the initial photon pairs to the photons received by Alice and Bob, as established by coincidences at the BSM. Each SNSPD in the setup detects approx. 0.5\,million QD photons per second.  To increase the entanglement swapping fidelity we use time gating of BSM detection events (gate width: $47$\,ps) at the expense of the total rate of heralding events. Quantum state tomography is performed using sets of four-fold coincidences at different polarization settings for Alice and Bob. The determined density matrix shown in Fig.~\ref{fig4} closely resembles the Bell state $\ket{\Psi^+}$. The fidelity of $f_{AB}=0.81\pm0.04$ clearly surpasses the classical limit of $0.5$ and therefore testifies to the successful swapping of the entangled state.

Fig.~\ref{fig4}c features the measurement of four-fold coincidences in the co- and cross-polarized diagonal bases as a function of the relative time delay between emission 1 and 2. In this fashion the temporal overlap of the XX photons at the beam splitter in the BSM is tuned. The highest XX photon indistinguishability is found at zero delay, resulting in a distinct coincidence offset for co- and cross-polarized bases. As the delay time departs from zero the BSM success starts to drop. This results in assimilating four-fold coincidences. The data, obtained without time gating at the BSM, can be well fitted to double-sided exponential functions denoted as solid lines.

The main bottleneck in reaching higher $f_{AB}$ is the XX photon indistinguishability. This can also be seen from $Re(\rho_{AB})$ in Fig.~\ref{fig4}b: The well-fitting diagonal elements are mainly determined by the high initial fidelities $f_1$ and $f_2$. However, the more deviant off-diagonal elements depend on both the degree of entanglement and the XX photon indistinguishability. In Fig.~\ref{fig4}d the fidelity $f_{AB}$ and the Bell parameter $S$, as used in the CHSH-Bell inequality~\cite{Bell1964,CHSH1969}, are plotted against the temporal gate width. For large gate widths the fidelity decreases to $f_{AB} = 0.71 \pm 0.03$. This is in perfect agreement with the calculated fidelity of $f_{AB} = 0.71$, given the observed entanglement fidelities $f_i$ and XX photon indistinguishability $I$ discussed above. The maximum achievable fidelity for our QD emission is $f_{max} = 0.89$, assuming unity indistinguishability (a gate width approaching zero). In reality this value cannot be approached due to the limited time resolution of the detectors. The Bell parameter $S=2.28 \pm 0.13$ at the $47$\,ps gate violates the CHSH-Bell inequality, $S\leq 2$, by more than two standard deviations. Assuming perfect indistinguishability it reaches $S_{max}=2.47$.\\

In a final step we investigate the compatibility of our semiconductor entangled photon source with atomic transitions of rubidium, a prominent quantum memory candidate. Maintaining entangled photon emission, the emission frequency is tuned over the Rb~$D_2$ transitions at $780.04$~nm by controlling the QD temperature~\cite{Ding2016}. Fig.~\ref{fig4}e displays the $^{87}$Rb vapor cell transmission against the relative frequency detuning of a spectrally narrow laser (black). Two prominent absorption features are observed corresponding to the two $^{87}$Rb ground states split by the hyperfine interaction~\cite{Penselin1962}. Residual $^{85}$Rb in the vapor cell results in the smaller absorption features visible at detunings of $-1$\,GHz and $2$\,GHz. The transmission of the QD photons (blue) shows two clear absorption dips, which are broadened due to the QD linewidth of $\Delta\nu = (4.9 \pm 0.2)$\,GHz. This opens the door for further experiments involving the storage of polarization-encoded qubits in atomic quantum memories. In addition, Rb atomic transitions could serve as a common and global reference at which the QD emission can be frequency-stabilized~\cite{Zopf2018}. Thus the indistinguishability of photons from distant nodes in a quantum network could be ensured.

\section{Discussion and Outlook}
Demonstrating entanglement swapping between photon pairs emitted from semiconductor QDs marks a milestone for quantum photonics, since these sources surpass existing technologies in terms of on-demand photon emission and scalability. Compatibility with atom-based quantum memories paves the way for hybrid quantum repeater implementations. An efficient photon-atom interface  requires the linewidths of both systems to be matched, e.g. by combining lifetime-limited QD emission and Purcell broadening of atomic lines~\cite{Gallego2018}.\\

Further experiments that now become feasible with these sources are entanglement swapping with photons from distant emitters, multi-photon entanglement or entanglement distillation. The outcome will be dictated by the optical quality of these sources. Promising improvements include silicon-integrated strain-tuning platforms, which facilitates the emission of wavelength-tunable entangled photons~\cite{Chen2016}. Integrating QDs into micro-cavities can increase their brightness and photon indistinguishability~\cite{Senellart2017,Sapienza2015}. Another key ingredient towards a scalable quantum photonic network is electrically triggered photon emission~\cite{Zhang2015a,Mueller2018}. Decoherence due to coupling to the solid-state environment can be controlled by electric fields in QD integrated diode structures~\cite{Kuhlmann2013}. Overcoming the challenge of combinig these techniques in fabricated devices will be a next major step in realizing semiconductor based quantum networks.

\section{Acknowledgments}
We acknowledge funding by the BMBF (Q.com) and the European Research Council (QD-NOMS). F.D. acknowledges support by IFW Excellence Program. We thank Wenjamin Rosenfeld (LMU M\"{u}nchen), Tobias Macha (Universit\"{a}t Bonn), Matthew Eiles (MPIPKS Dresden), and Franz L\"{o}chner (FSU Jena) for fruitful discussions.

\section{Author contributions}
The experiment was conceived by F.D. and O.G.S. who directed the research. The quantum dot samples were grown by R.K and integrated with the broadband optical antenna by Y.C. The optical setup was built and tested by M.Z., J.Y. and D.C. Optical measurements and data analysis were performed by M.Z. with help from J.Z. and R.K. The manuscript was written by M.Z., with input from all the authors.



\end{document}